# The impact of Ozone on Earth-like exoplanet climate dynamics: the case of Proxima Centauri b

P. De Luca[1], M. Braam[2,3,4], T.D. Komacek[5] and A. Hochman[6]

1. Barcelona Supercomputing Center (BSC), Barcelona, Spain
2. School of GeoSciences, The University of Edinburgh, Edinburgh, UK
3. Centre for Exoplanet Science, The University of Edinburgh, Edinburgh, UK
4. Institute of Astronomy, KU Leuven, Leuven, Belgium
5. Department of Astronomy, University of Maryland, College Park, USA
6. Fredy and Nadine Hermann Institute of Earth Sciences, The Hebrew University of Jerusalem, Jerusalem, Israel

Corresponding author: Paolo De Luca, paolo.deluca@bsc.es

## Abstract

The emergence of the James Webb Space Telescope and the development of other advanced observatories (e.g., ELTs, LIFE and HWO) marks a pivotal moment in the quest to characterize the atmospheres of Earth-like exoplanets. Motivated by these advancements, we conduct theoretical explorations of exoplanetary atmospheres, focusing on refining our understanding of planetary climate and habitability. Our study investigates the impact of ozone on the atmosphere of Proxima Centauri b in a synchronous orbit, utilizing coupled climate chemistry model simulations and dynamical systems theory. The latter quantifies compound dynamical metrics in phase space through the inverse of co-persistence ($\theta$) and co-dimension ($d$), of which low values correspond to stable atmospheric states. Initially, we scrutinized the influence of ozone on temperature and wind speed. Including interactive ozone (i.e., coupled atmospheric (photo)chemistry) reduces the hemispheric difference in temperature from 68 °K to 64 °K, increases (~+7 °K) atmospheric temperature at an altitude range of ~20–50 km, and increases variability in the compound dynamics of temperature and wind speed. Moreover, with interactive ozone, wind speed during highly temporally stable states is weaker than for unstable ones, and ozone transport to the nightside gyres during unstable states is enhanced compared to stable ones (~+800 DU). We conclude that including interactive ozone significantly influences Earth-like exoplanets' chemistry and climate dynamics. This study establishes a novel pathway for comprehending the influence of photochemical species on the climate dynamics of potentially habitable Earth-like exoplanets. We envisage an extension of this framework to other exoplanets.

## 1. Introduction

The recent discovery of many nearby potentially temperate terrestrial exoplanets provides the opportunity to characterize climates of planets that may be Earth-like (Anglada-Escudé et al., 2016; Delrez et al., 2022; Gillon et al., 2017; Kossakowski et al., 2023; Rodriguez et al., 2020). Importantly, these temperate planets all have close-in orbits around M-dwarf host stars, which is expected to cause their basic-state climate to be impacted by tidal spin-synchronization (Pierrehumbert & Hammond, 2019). Most notably, the large day-to-night irradiation contrast is expected to drive strong dayside convection (Sergeev et al., 2020; Yang et al., 2013) and a planetary-scale Matsuno-Gill pattern of equatorial waves (Showman et al., 2013). The combination of dayside convection, potential equatorial super-rotation, and off-equatorial Rossby waves are expected to control the spatial distribution of atmospheric tracers, including clouds (Komacek & Abbot, 2019; Suissa et al., 2020) and photochemically produced chemical species such as ozone (Braam et al., 2023; Chen et al., 2021).

Ozone ($O_3$) forms from the photolysis of molecular oxygen ($O_2$) via the Chapman mechanism (Chapman, 1930). On Earth, atmospheric $O_2$ is produced by photosynthesis and thus an indication of life. Therefore, $O_2$ and its photochemical by-product ozone have been proposed as potential biosignatures (e.g., Des Marais et al., 2002; Schwieterman et al., 2018). However, a variety of exoplanet scenarios, including the different ratios of near-UV to far-UV fluxes that planets around M-dwarfs receive, may drive abiotic build-up of $O_2$ and ozone in planetary atmospheres (e.g., Domagal-Goldman et al., 2014; Selsis et al., 2002).

Ozone is also a radiatively active species and thus impacts a planetary atmosphere's thermal and dynamic structure. Being both (photo)chemically and radiatively active makes ozone one of many potential species that induce climate-chemistry interactions, which have received considerable attention in models of the Earth System (e.g., Isaksen et al., 2009; Ramanathan et al., 1987), also as part of the Inter-governmental Panel on Climate Change reports (IPCC, 2021). The radiative impact of ozone consists of two components. First, the stratospheric ozone layer absorbs incoming UV radiation on Earth. This layer protects the surface from harmful UV fluxes, which is also the case when considering the radiation received by exoplanets around M, K, G, and F-type stars (e.g., Segura et al., 2003, 2005) or the 3-D impact of stellar flares from M-dwarfs (Ridgway et al., 2023). Ozone then emits thermal energy at infrared wavelengths, heating the stratosphere and producing a temperature inversion, which is also predicted for various exoplanets (e.g., Godolt et al., 2015). Second, ozone in the upper troposphere acts as a greenhouse gas, absorbing the outgoing radiation from a planet. The radiative effect of ozone strongly depends on the 3-D (vertical and horizontal) distribution through the atmosphere, which is ultimately determined by the complex interplay between (photo)chemistry and atmospheric dynamics. Studying such climate-chemistry interactions motivates using a 3-D coupled climate-chemistry model (CCM).

A growing body of work uses CCM simulations to study Earth-like exoplanets in a synchronized orbit. Generally, significant hemispheric contrasts in ozone exist for planets around M-dwarfs (e.g., Chen et al., 2018; Yates et al., 2020), and such spatial variations should affect future observations (Cooke et al., 2023). Simulations of Proxima Centauri b predict a significant zonal structure in the ozone distribution with accumulation of ozone at the gyre locations (Yates et al., 2020), driven by a stratospheric circulation connecting the photochemically active dayside and the gyres on the nightside (Braam et al., 2023). Furthermore, Yates et al. (2020) have shown that steady-state climate conditions differ when ozone is computed interactively, depending on photochemistry, atmospheric circulation, and temperature, compared to the simulations without ozone and with a fixed Earth-like ozone profile (Boutle et al., 2017). However, simulations of Proxima Centauri b also show different versions of internal variability in, for example, stratospheric winds (Cohen et al., 2022) and planetary-scale waves (Cohen et al., 2023), affecting the climate and distributions of atmospheric tracers. At present, a full assessment of the impact of ozone on key atmospheric variables and climate dynamics by also applying a novel dynamical systems theory approach is the logical next step (Faranda et al., 2020; De Luca, Messori, Faranda, et al., 2020; De Luca, Messori, Pons, et al., 2020).

Dynamical systems theory is the discipline that studies the trajectory of a given chaotic dynamical system, such as an atmosphere, by analyzing Poincaré recurrences in phase space. Recent developments have seen Poincaré recurrences combined with extreme value theory (Faranda et al., 2017; Lucarini et al., 2012). This approach allows us to quantify two main metrics, the inverse of local persistence ($\theta$) and local dimension ($d$), which are instantaneous in time and computed for 2-D latitude-longitude maps. The former indicates the mean residence time of a given number of states around a state of interest. The latter provides information about the active degrees of freedom of the same states in the atmospheric phase space. The lower the $\theta$ and $d$, the more stable (in the sense of atmospheric variability) the atmospheric motion of the variable of interest, whereas the higher the $\theta$ and $d$, the less stable the trajectory of the atmospheric variable. The methodology has been used for univariate atmospheric variables such as temperature, precipitation, and geopotential height, and various studies proved its usefulness in analyzing the climate dynamics of Earth (e.g., Hochman et al., 2019, 2020;

Vakrat & Hochman, 2023; Wedler et al., 2023) and terrestrial exoplanets (Hochman et al., 2022, 2023). The method has also more recently been expanded to assess two (or more) atmospheric variables simultaneously (Faranda et al., 2020; De Luca, Messori, Faranda, et al., 2020; De Luca, Messori, Pons, et al., 2020). Moreover, in this compound case, one can obtain two dynamical metrics: the inverse of local co-persistence ($\theta_{T, WS}$) and local co-dimension ($d_{T, WS}$). They resemble the univariate metrics, with the only difference being that they are computed from two atmospheric variables of interest, temperature (T) and wind speed (WS), in two different phase spaces. Therefore, their values are obtained from joint recurrences, and we define them here as compound dynamical metrics.

Our main aim is to leverage climate model simulations of Proxima Centauri b, traditional atmospheric analysis, and dynamical systems theory to understand how ozone influences key atmospheric variables and climate dynamics. Section 2 describes the climate model setup and data, the computation of the compound dynamical systems metrics, and statistical inference. In Section 3, we present our results regarding the direct effects of ozone on atmospheric variables and dynamics. We discuss our findings for Proxima Centauri b and provide conclusions extended to the broad range of Earth-like exoplanets in Section 4.

## 2. Methods

### 2.1 Coupled climate-chemistry model

We use a 3-D coupled Climate-Chemistry Model (CCM) consisting of the Met Office Unified Model (UM) and the UK Chemistry and Aerosol framework (UKCA). Braam et al. (2022) described the CCM in extensive detail. Here, we limit ourselves to a brief description of the essential components relevant to this study. A 3-D CCM simulates the coupled evolution of radiative transfer, dynamics, and chemistry in a planetary atmosphere, comprehensively assessing the planetary climate and habitability.

The UM is a versatile General Circulation Model, using the ENDGame dynamical core to solve the equations of motion (Wood et al., 2014) and the Suite of Community Radiative Transfer codes based on the Edwards and Slingo scheme to compute radiative transfer (SOCRATES; Edwards & Slingo, 1996). The incoming stellar radiation from Proxima Centauri (M5.5V star) is based on the v2.2 composite spectrum from the MUSCLES spectral survey (France et al., 2016; P. Loyd et al., 2016; Youngblood et al., 2016). Sub-grid scale processes like convection (Gregory & Rowntree, 1990), water cloud physics (Wilson et al., 2008), and turbulent mixing (Brown et al., 2008; Lock et al., 2000) are parametrized. Besides the common use of the UM in predicting the Earth's weather and climate, it has been adapted to Mars (McCulloch et al., 2023) and exoplanets ranging from terrestrial planets (e.g., Boutle et al., 2017; Mayne, Baraffe, Acreman, Smith, Wood, et al., 2014) to hot Jupiters (e.g. Amundsen et al., 2016; Mayne, Baraffe, Acreman, Smith, Browning, et al., 2014). UKCA is a framework to simulate 3-D kinetic and photo-chemistry (Archibald et al., 2020; Braam et al., 2022; Morgenstern et al., 2009; O'Connor et al., 2014). The atmospheric transport of chemical tracers is fully coupled to the UM's large-scale advection, convection, and boundary layer mixing. Here, we limit ourselves to UKCA's gas-phase (photo)chemistry description. The chemical network consists of 21 chemical species connected by 71 reactions (as shown in the appendix of Braam et al., 2022). The network describes the Chapman mechanism of ozone formation as well as the catalytic destruction of ozone by the hydrogen oxide and nitrogen oxide catalytic cycles, representing interactive ozone chemistry in the 3-D model.

|  | No-Chemistry | Chemistry |
|---|---|---|
| Semimajor axis (au) | 0.0485 | |
| Stellar irradiance (W m$^{-2}$) | 881.7 | |
| Orbital period (d) | 11.186 | |
| Rotation rate (rad s$^{-1}$) | $6.501 \times 10^{-6}$ | |
| Radius (R$_\oplus$) | 1.1 | |
| Surface gravity (m s$^{-2}$) | 10.9 | |
| Surface pressure (bar) | 1 | |
| CO$_2$ mass fraction (kg kg$^{-1}$) | $5.941 \times 10^{-4}$ | |
| H$_2$O mass fraction (kg kg$^{-1}$) | $1.00 \times 10^{-8} - 1.33 \times 10^{-3}$ | $1.28 \times 10^{-8} - 1.29 \times 10^{-3}$ |
| O$_2$ mass fraction (kg kg$^{-1}$) | 0 | 0.2314 |
| O$_3$ mass fraction (kg kg$^{-1}$) | 0 | $3.29 \times 10^{-8} - 5.46 \times 10^{-5}$ |

**Table 1** - Orbital, planetary, and atmospheric parameters used to configure Proxima Centauri b following Boutle et al. (2017). Note that the mass fractions of $H_2O$ and $O_3$ are interactively calculated; therefore, they are given as a range indicating minimum and maximum values. $H_2O$ follows from the balance of evaporation, condensation, (photo)chemistry, and $O_3$ from (photo)chemistry.

We simulate Proxima Centauri b (Anglada-Escudé et al., 2016) as an aqua planet orbiting in a 1:1 spin-orbit resonance around its M-type host star, using the orbital parameters described in Table 1. Anglada-Escudé et al. (2016) well-constrained the semimajor axis and orbital period. Given the close-in orbit at 0.0485 AU, a 1:1 spin-orbit resonance or synchronous orbit is a probable scenario (Barnes, 2017) and results in a rotation rate of $6.501 \times 10^{-6}$ rad s$^{-1}$. The stellar irradiance follows from Boutle et al. (2017). Since Proxima Centauri b is non-transiting, we only have a lower limit on its mass from the detection by the radial velocity method of $M_p \sin(i) = 1.07 \pm 0.06$ M$_\oplus$ (Anglada-Escudé et al., 2016; Faria et al., 2022), where $i$ is the orbital inclination. Given the unknown actual mass of Proxima Centauri b and to ensure consistency with previous GCM simulations, we follow Turbet et al. (2016) and assume an actual planet mass of 1.4 M$_\oplus$. Assuming that Proxima Centauri b has Earth's density (5.5 g cm$^{-3}$), we estimate a corresponding planetary radius of 1.1 R$_\oplus$ and a surface gravity of 10.9 m s$^{-2}$. While these configurations are based on the known parameters of Proxima Centauri b, the simulation results can apply more generally to planets with similar sizes and rotation periods in spin-synchronous orbits around M-dwarf stars. We use a horizontal resolution of $2° \times 2.5°$ in latitude and longitude, respectively, with the substellar point at $0°$ latitude and longitude. We assume the entire surface is covered by a 2.4 m slab ocean mixed layer with a total heat capacity of $10^7$ J K$^{-1}$ m$^{-2}$. We simulate an atmosphere extending up to 85 km altitude, divided over 60 vertical levels that are quadratically stretched for enhanced near-surface resolution (Yates et al., 2020). Abundances of $N_2$, $O_2$, and $CO_2$ (Table 1) correspond to pre-industrial Earth levels, and water vapor profiles are determined interactively following evaporation from the slab ocean.

We initialize the No-Chemistry setup with the uniform mass mixing ratios of $N_2$, $CO_2$, and $H_2O$ from surface evaporation. $CO_2$ and $H_2O$, as radiatively active species, are important factors in the thermodynamic and dynamic state of the atmosphere. For the Chemistry simulation, we also specified a uniform mass mixing ratio for $O_2$ and included the interactive calculation of ozone chemistry. Ozone is another radiatively active species, and its interactively determined and varying mixing ratios are used in the radiative transfer calculations, potentially affecting the thermodynamic and dynamic state of the atmosphere. We spin up both simulations for ~20 Earth years to ensure a steady state, diagnosed by radiative balance, surface temperatures, and ozone abundances. After that, we run the simulations for another 30 years with daily output to analyze Proxima Centauri b's atmospheric dynamics.

We extract two atmospheric variables from the 30-year Chemistry and No-Chemistry simulations: temperature (T in °K), linked with thermodynamic processes, and wind speed (WS in m s$^{-1}$), related to dynamic processes. In addition, we obtain the vertically integrated ozone column density in DU (1 DU = 2.69·10$^{20}$ molecules m$^{-2}$) and ozone mass fraction (kg kg$^{-1}$) over the same levels, hereafter OzCol and OzFr, respectively. From these four variables, we use all 60 vertical levels when assessing their vertical profiles and choose one level for computing composite and difference maps. This level corresponds to ~22km for T, WS, and OzFr since we find the largest concentration of ozone particles at this level. For OzCol, we use the surface level representing the vertically integrated amount of overhead ozone molecules in the vertical column. We then provide vertical profiles for these four atmospheric variables and compute them for the global, northern, and southern gyre regions. For each variable, simulation, and atmospheric vertical level, we take the temporal median over the entire 30-year period and then compute the field median. This leaves us with 60 data points for each variable and region representing the atmosphere.

## 2.2 Dynamical Systems Metrics

We used a novel dynamical systems method to compute two compound dynamical metrics: the inverse of local co-persistence and local co-dimension, which we refer to as $\theta_{T, WS}$ and $d_{T, WS}$, respectively. The metric $\theta_{T, WS}$ is intuitively a measure of the joint average residence time of two trajectories around two respective states of interest. The lower the value of $\theta_{T, WS}$, the more likely it is that the preceding and future states of the systems will resemble the current states. The metric $d_{T, WS}$ describes the joint evolution of the systems around two respective states of interest and can be interpreted as a proxy for the joint number of degrees of freedom active around the same states (Faranda et al., 2020; De Luca, Messori, Faranda, et al., 2020; De Luca, Messori, Pons, et al., 2020).

Calculating the compound dynamical systems metrics combines Poincaré recurrences with extreme value theory (Faranda et al., 2017, 2020; Freitas et al., 2010; Lucarini et al., 2012). We referred to the system returning *n* times close to a previously visited state in the phase space as *recurrences*. We considered an atmospheric variable, such as T, and a given state of interest, namely a 2-D map of a day within the time series of T. Our approach uses the Euclidean distance to quantify how close the state of interest and the recurrences are to one another in the atmospheric phase space (Faranda et al., 2017).

We considered a second phase space for the latter variable to extend the dynamical systems analysis to two variables, T and WS. Eventually, we computed joint recurrences around a common state of interest, corresponding to two instantaneous latitude–longitude maps: one for T and one for WS. Once we defined the joint recurrences, we computed $\theta_{T,\,WS}$ and $d_{T,\,WS}$ compound metrics for the 60 vertical levels in the Proxima Centauri b's climate model simulations. The final output of such analysis is a value for each metric, daily time step, and vertical level. This allows us to relate specific metric values to the corresponding geographical patterns of selected atmospheric variables. We further computed the compound dynamical metrics for particular regions of interest: the northern gyre, where atmospheric ozone tends to accumulate, and equatorial western and eastern terminators, which are highly variable (Figure 1e). Finally, we defined 'High' and 'Low' dynamically stable days as the lower and upper 5% of $\theta_{T,\,WS}$ and $d_{T,\,WS}$ compound metrics, respectively. For a complete derivation of $\theta_{T,\,WS}$ and $d_{T,\,WS}$ we refer the reader to Faranda et al. (2020).

### 2.3 Statistical inference

We perform a two-tailed Wilcoxon rank-sum test to assess the statistical significance of the median vertical profiles and difference maps (Mann & Whitney, 1947). The test was performed between the Chemistry and No-Chemistry medians, composite datasets, and between compound stable and unstable atmospheric states under the null hypothesis that the medians of the datasets are equal. We then compute the composite maps' corresponding p-values at the grid-point level. To account for Type I errors (or false positives), we apply the Bonferroni correction of p-values (Bonferroni, 1936), which divides the original p-values by the total number of statistical tests performed. Lastly, to assess the statistical significance of the standard deviation's vertical profiles, we perform the F-test under the null hypothesis that the variances of the two populations are equal (Snedecor & Cochran, 1989). We provide all statistical tests at the 1% significance level.

### 3. Results

### 3.1 How does interactive Ozone influence the climatology of atmospheric variables?

Figure 1 shows the 30-year climatology of four atmospheric variables, i.e., T, WS, OzCol, and OzFr, for both Chemistry and No-Chemistry simulations. We note that the stratospheric temperature for the Proxima Centauri b Chemistry simulation increases globally compared to the No-Chemistry due to the different chemical compositions (Figure 1a-b). The ozone layer (Figure 1e) and particularly stratospheric ozone (Figure 1f) in the Chemistry simulation absorbs incoming ultraviolet radiation and reemits this at infrared wavelengths, leading to warming in the stratosphere. On the other hand, the temperature gradient between the two simulations is kept, meaning that T at higher (northern and southern) latitudes reaches its minimum and gradually increases meridionally towards the equatorial region (Figure 1a-b). The wind speed shows a similar spatial distribution for the Chemistry and No-Chemistry climatology (Figure 1c-d), with lower wind speed at higher latitudes and an increase towards the equatorial jet. Nevertheless, the jet's intensity changes with weaker eastward winds at the equator for the Chemistry simulation. Taking the surface temperature as a proxy for day-to-night contrasts, we determine a hemispheric contrast (dayside average minus nightside average) of 64 °K and 68 °K for the Chemistry and No-Chemistry climatology, respectively. The smaller day-to-night temperature contrast for the Chemistry climatology may be the reason for a weaker jet considering thermal wind balance. Lastly, both ozone climatologies show similar patterns with the higher concentration of ozone in the northern and southern gyre regions (Figure 1e-f). In contrast, for OzFr, we still observe higher ozone values in the proximity of the gyres and across the mid and higher latitudes with a meridional gradient of lower values towards the equator (Figure 1f). The difference between OzCol and OzFr climatology is due to the former representing ozone integrated over the entire vertical column, whereas the latter is strictly confined to stratospheric ozone fraction. Spatial variations in the distribution of ozone are driven by stratospheric circulation mechanisms, including an analogue of the Brewer-Dobson circulation that controls the ozone distribution on Earth and the stratospheric dayside-to-nightside circulation for synchronously rotating planets (Braam et al., 2023).

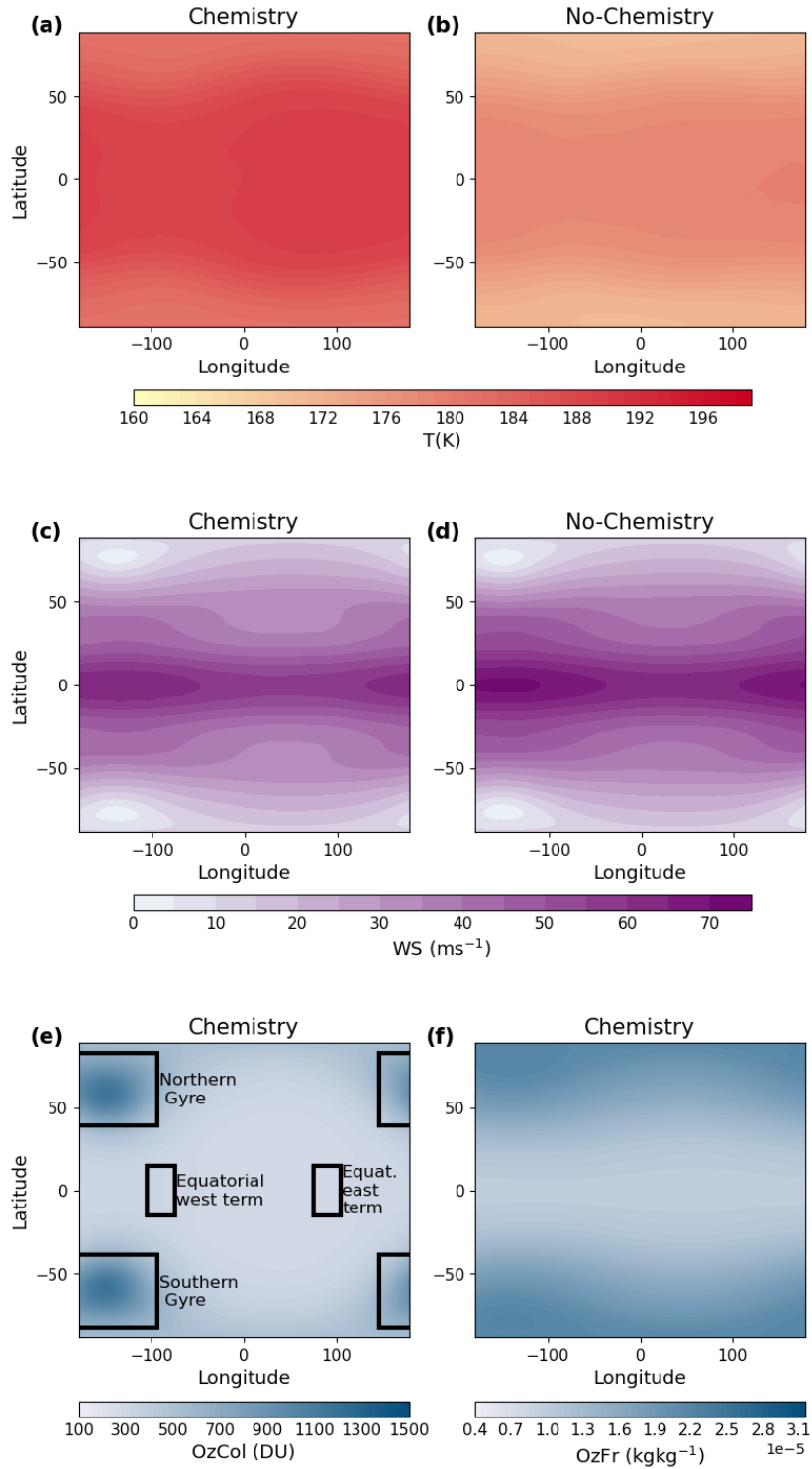

**Figure 1** - Climatology for key variables in the Chemistry and No-chemistry model simulations. (a-b) Temperature (T in °K). (c-d) Wind speed (WS in m s$^{-1}$). (e) Ozone column density (OzCol in DU). (f) Ozone fraction (OzFr in kg kg$^{-1}$). Median composites were computed over the 30 years for Chemistry (a, c, e, f) and No-Chemistry (b, d) simulations. Climatology of (a-d, f) was computed from the ~22km level, whereas for (e) from the surface level. In (e) we mark the geographical regions used with rectangles: northern gyre (Lon 146.25°, -93.25°, Lat 39°, 83°); southern gyre (Lon 146.25°, -93.25°, Lat -39°, -83°); equatorial western terminator (Lon -105°, -75°, Lat -15°, 15°); and equatorial eastern terminator (Lon 75°, 105°, Lat -15°, 15°).

The temperature vertical profiles look similar between the global and gyre regions (Figure 2a-c). However, both gyres exhibit lower temperatures (~175 °K) than the global median (~211 °K) at the surface. Such a temperature difference is noticeable within the first 5-6 km from the surface because these gyres trap air, subject to extensive radiative cooling due to the nightside location. For both Chemistry and No-Chemistry and all three regions, we observe an increase in temperature, then an inversion, and a slight increase up to the top of the atmosphere. A significant difference is that the temperature for the Chemistry simulation is significantly higher (~+7 °K) than the No-Chemistry one from ~17 km to ~50 km in all three regions, probably because at this altitude range, we find the highest ozone mass fraction (Figure 2j-l). The vertical profiles for wind speed are also very similar between all the regions in the Chemistry and No-Chemistry simulations (Figure 2d-f). However, the wind speed in the Chemistry simulation in the gyres is significantly higher (~+5 m s$^{-1}$) than the No-Chemistry one. This may be due to multiple complex factors since the location of the gyres and strength of the rotating winds will depend on the radiative forcing and heat transport (e.g., Pierrehumbert & Hammond, 2019; Showman et al., 2013), which slightly change due to the inclusion of ozone. The values of OzCol are higher from ~0 km to ~20 km over the gyres compared to the global region (Figure 2g-i; Braam et al., 2023). The OzCol = 0 DU from ~40 km upwards for all three regions. However, OzFr in all three regions increases from the surface to ~48 km when it reaches its maximum and then decreases close to 1e-08 kg kg$^{-1}$ at ~76 km (Figure 2j-l). Between the global and the gyre profiles, we notice a difference close to the peak in OzFr: the vertical distribution of ozone over the gyre regions shows a saddle from ~48 km to ~60 km due to the formation of a secondary ozone layer on the nightside hemisphere in the absence of ozone photolysis (e.g., Smith & Marsh, 2005).

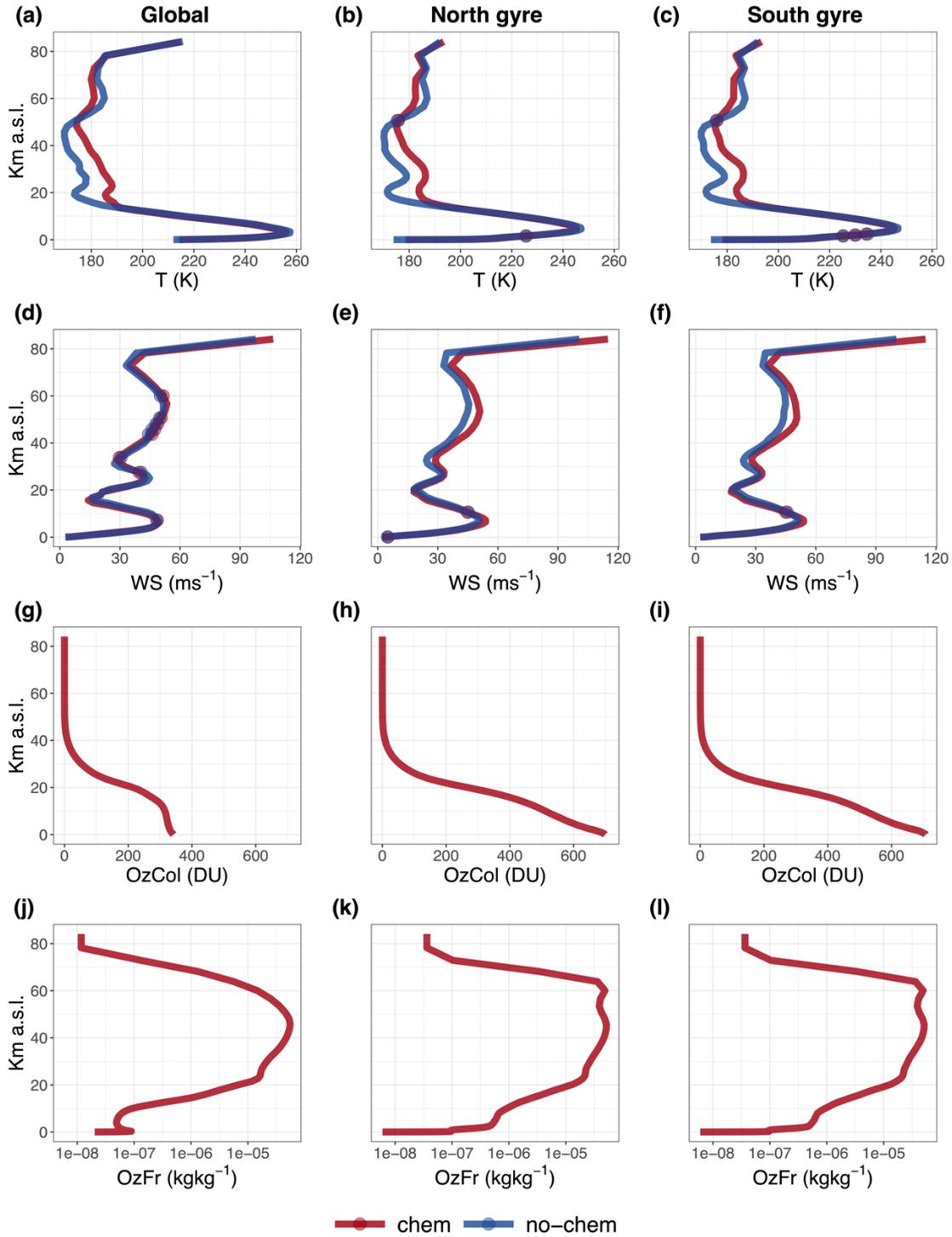

**Figure 2** - Vertical profiles of four atmospheric variables. (a-c) Temperature – T, (d-f) Wind speed – WS, (g-i) Ozone column – OzCol and (j-l) Ozone fraction – OzFr. The first column represents vertical profiles computed from global medians over the 30 years. In contrast, the second and third columns represent the same, but for the northern and southern gyres, where the concentration of ozone is higher. (a-f) Profiles of the Chemistry and No-Chemistry simulations. (g-l) Profiles computed from the Chemistry simulation. Circles in (a-f) represent Chemistry and No-chemistry medians that are *not significantly* different at the 1% level. OzFr in (j-l) is plotted on a $\log_{10}$ scale.

## 3.2 How does interactive ozone influence the atmospheric dynamics?

Here, we analyze the influence that interactive ozone has on the dynamics of Proxima Centauri b's atmosphere. In this respect, we show the median vertical profiles of the compound dynamical systems metrics (Figure 3). Globally, although relatively small, we find significant differences between the Chemistry and No-Chemistry simulations. A substantial decrease in $\theta_{T, WS}$ and $d_{T, WS}$ is observed below the levels of maximum OzFr and an increase above these levels (compare Figure 3a, e with Figure 2j). Some levels, especially for $d_{T, WS}$, do not show significant differences. These findings are also apparent in the north gyre (Figure 3b, f) and west and east terminators (Figure 3c-d, g-h). However, for the terminators, we note that Chemistry's and No-Chemistry's *co-dimension* shows lower values over the entire vertical profile when compared to the global and north gyre regions. We further display the standard deviations vertical profiles of $\theta_{T, WS}$ and $d_{T, WS}$ (Figure 3i-p). We provide evidence for significantly larger variability in the atmospheric time series dynamics of Proxima Centauri b when including interactive ozone than not having it. This finding is particularly evident just below the level of maximum ozone accumulation.

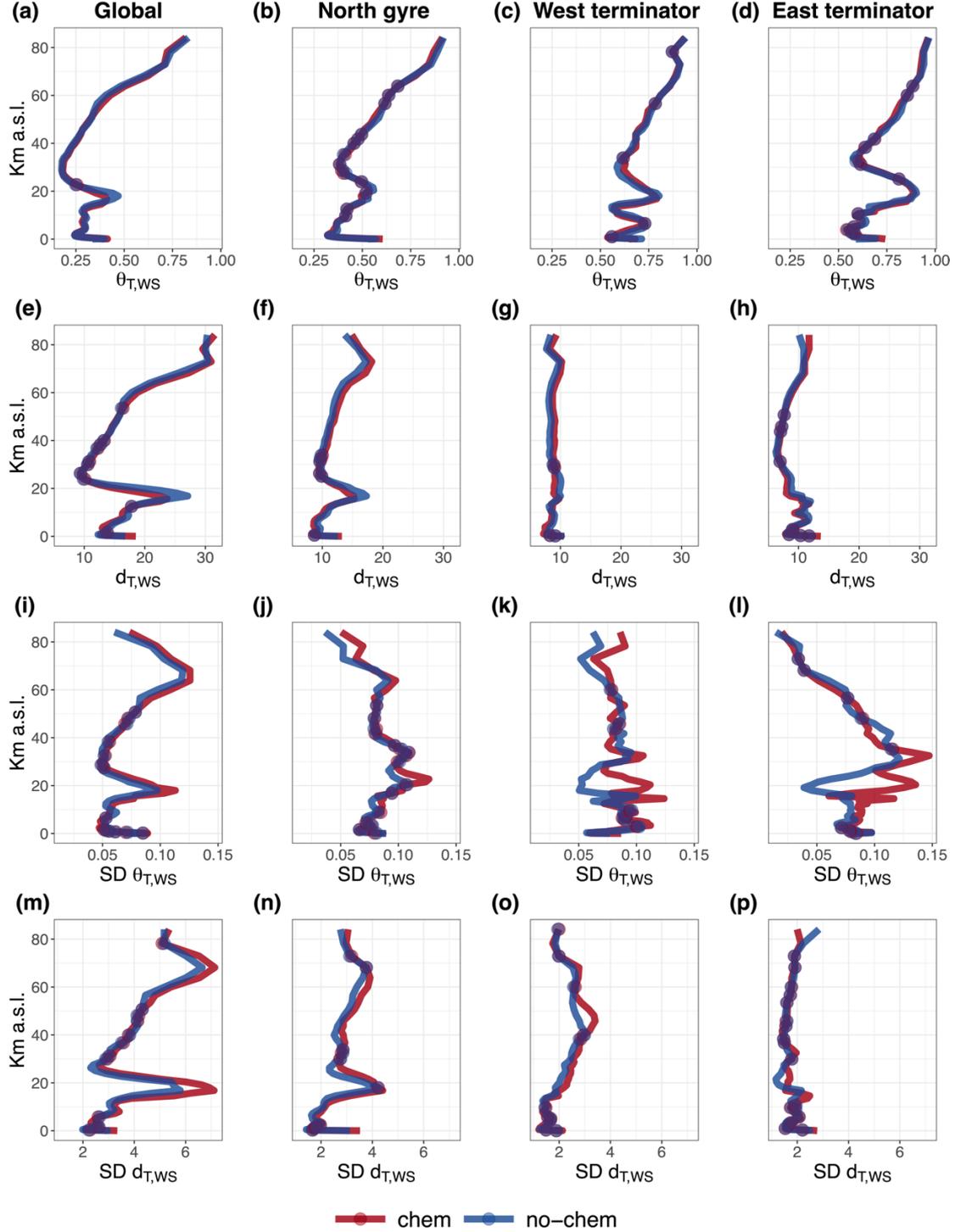

**Figure 3** - Vertical profiles of compound dynamical systems metrics. (a-d) The inverse of local co-persistence ($\theta_{T,\ WS}$). (e-h) Local co-dimension ($d_{T,\ WS}$) medians for Chemistry and No-Chemistry simulations. (i-p) the same but for the standard deviations. The first column shows the global vertical level values over the 30 years, the same for the second column but for the northern gyre. The third and fourth columns are for the western and eastern equatorial terminators. The dynamical system metrics are computed from Temperature (T) and Wind Speed (WS). Circles represent (a-h) medians and (i-p) standard deviations that are *not significantly* different at the 1% level.

Next, we selected 'High' (upper 5%) and 'Low' (lower 5%) $\theta_{T,WS}$ and $d_{T,WS}$ days in the Chemistry and No-Chemistry simulations (Figure 4). Joint low values reflect more stable (in the sense of atmospheric time series variability) atmospheric states, whereas joint higher values are relatively unstable atmospheric configurations. Also, from Figure 4, it is possible to notice an increased variability for the Chemistry simulation compared to the No-Chemistry since the data points in the former tend to spread more over the x and y axis compared to the latter.

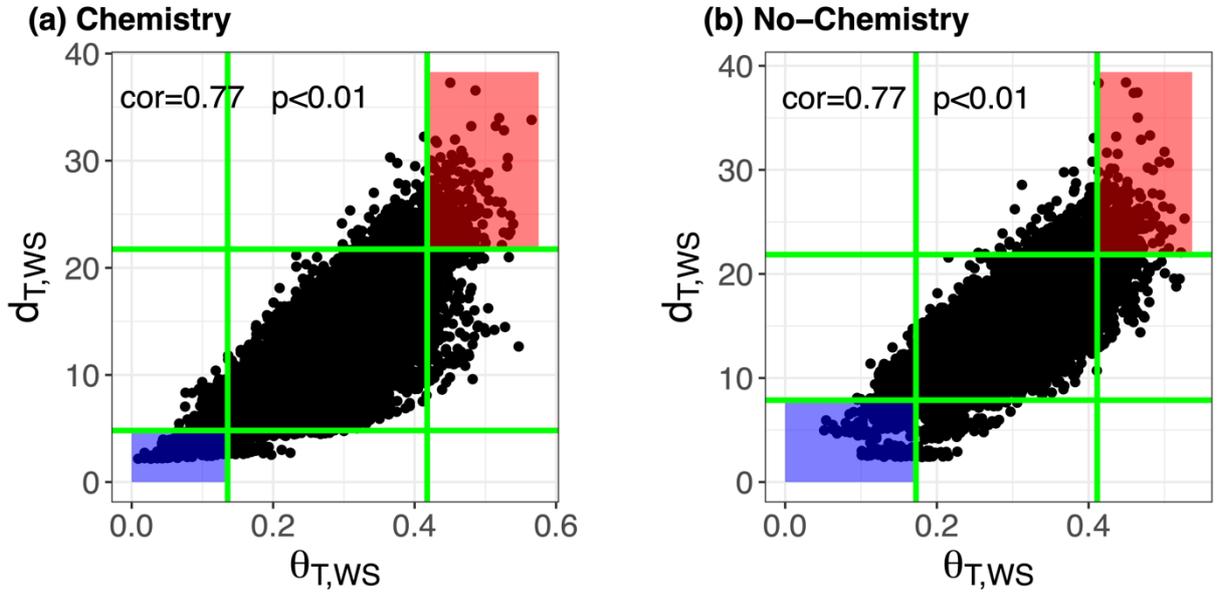

**Figure 4** - Scatter plots of $\theta_{T,WS}$ (x-axis) and $d_{T,WS}$ (y-axis) for (a) Chemistry and (b) No-Chemistry simulations. The compound dynamical systems metrics are computed from temperature and wind speed at the ~22km level. Blue and red shaded areas contain states with $\theta_{T,WS}$ and $d_{T,WS}$ <5th and >95th percentiles, 'Low' and 'High', respectively. On the top left of each panel, we show the Spearman's correlation coefficient and p-value.

We calculate field medians or composite maps for the 'High' and 'Low' atmospheric states from Figure 4. Next, we analyzed the differences between the median composite maps of temperature and wind speed for 'High' and 'Low' atmospheric states in the Chemistry and No-chemistry simulations at the vertical level corresponding to ~22 km altitude (Figures 5-6). Figure 5 shows the composite and difference maps for temperature. Here, the Chemistry simulation has higher temperatures in both 'High' and 'Low' states than the No-Chemistry simulation, thanks to the large abundance of OzFr at this level leading to radiative heating. In addition, temperature patterns are symmetrical between the northern and southern regions of Proxima Centauri b, with lower values at higher latitudes and higher values across the tropical region (Figure 5a-d). Difference maps of High - Low states point toward enhanced temperatures for the 'High' states over the planet, with both Chemistry and No-Chemistry simulations also showing slightly higher values in the gyre regions (up to +6 °K). This pattern is more pronounced in the former simulation (Figure 5e-f), and this is due to a higher accumulation of ozone in the 'High' states compared to 'Low' ones. Difference maps for Chemistry - No-Chemistry shows positive and significant temperature differences over the entire planet for both 'High' and 'Low' states (up to +15 °K) again due to ozone's radiative heating. Nevertheless, we note that 'High' states' positive temperature differences are observed over the midlatitudes, whereas the 'Low' states positive temperature differences occur over the tropics (Figure 5g-h).

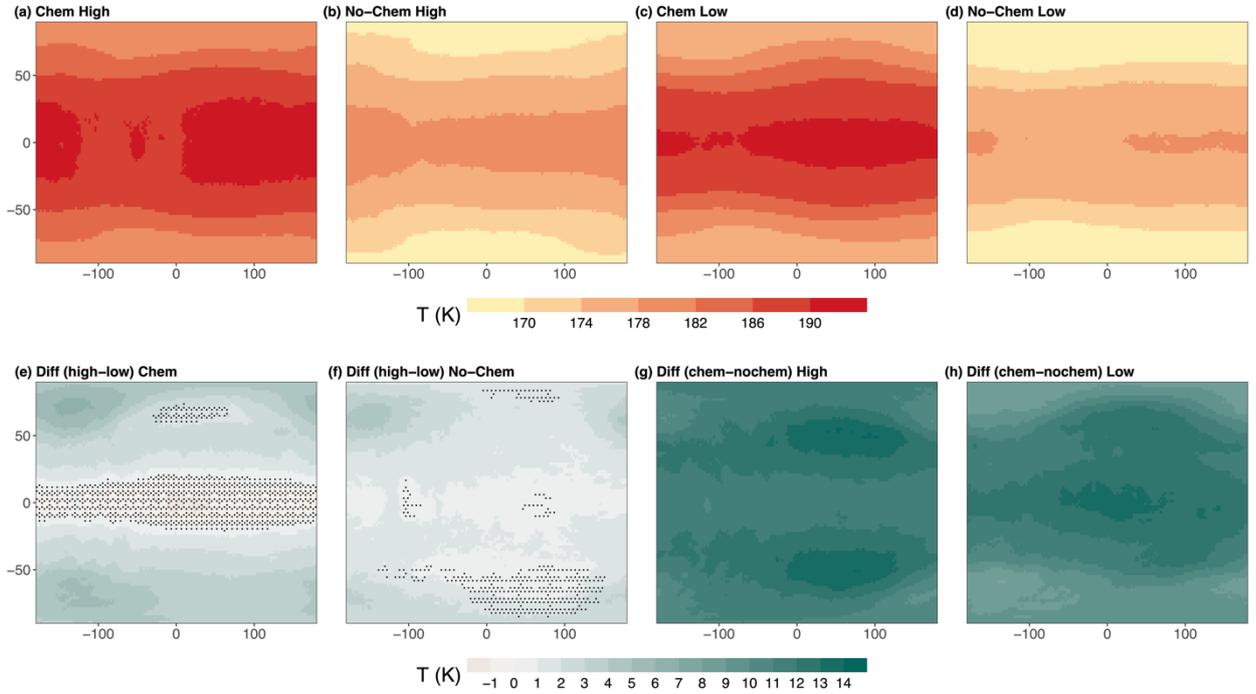

**Figure 5** - Composite and difference maps for Temperature (T) at the ~22km level. (a-b) Composite maps are field medians computed from Chemistry and No-Chemistry joint *co-persistence* and *co-dimension* states (defined in Figure 4), which are >95th percentile – 'High,' and (c-d) the same but for states <5th percentile – 'Low.' (e-f) Difference maps were computed by subtracting the 'Low' from the 'High' composites for Chemistry and No-Chemistry simulations. (g-h) Difference maps were calculated by subtracting the No-Chemistry to Chemistry composites for both 'High' and 'Low' states. In (e-h), stippling represents areas that are *not significantly* different at the 1% level.

Composite maps for wind speed show very similar and symmetrical patterns of high wind speed over the tropics (corresponding to the equatorial jet) and lower wind speed in the higher latitudes, with the only difference being the Chemistry 'Low' states simulation that has weaker wind speed values in the tropics (Figure 6a-d). This shows that the addition of interactive ozone substantially affects the persistent atmospheric states by weakening the equatorial jet (Figure 6c). This may imply that the mechanism that drives the equatorial jet on synchronously rotating exoplanets (Showman et al., 2013) diminishes in strength with the inclusion of ozone. This causes less pronounced gyres, which we will show with the composite maps of the OzCol below. In Section 4, we will put these findings on the equatorial jet and gyres into the context of the large-scale atmospheric circulation. Difference maps between High - Low show positive and high wind speed differences (up to +35 m s$^{-1}$) from ~50°S to ~50°N, with higher values for the Chemistry simulation than the No-Chemistry one. In the Chemistry simulation, we also provide evidence for negative wind speed differences (up to -10 m s$^{-1}$) over the gyres (Figure 6e-f). Difference maps for Chemistry - No-Chemistry 'High' states show positive wind speed values (up to +15 m s$^{-1}$) over most of the planet and negative differences in the gyres (up to -5 m s$^{-1}$). The same maps for 'Low' states show stronger negative wind speed differences (up to -25 m s$^{-1}$) from ~50°S to ~50°N, and weak positive differences for most of the remaining planet (Figure 6g-h).

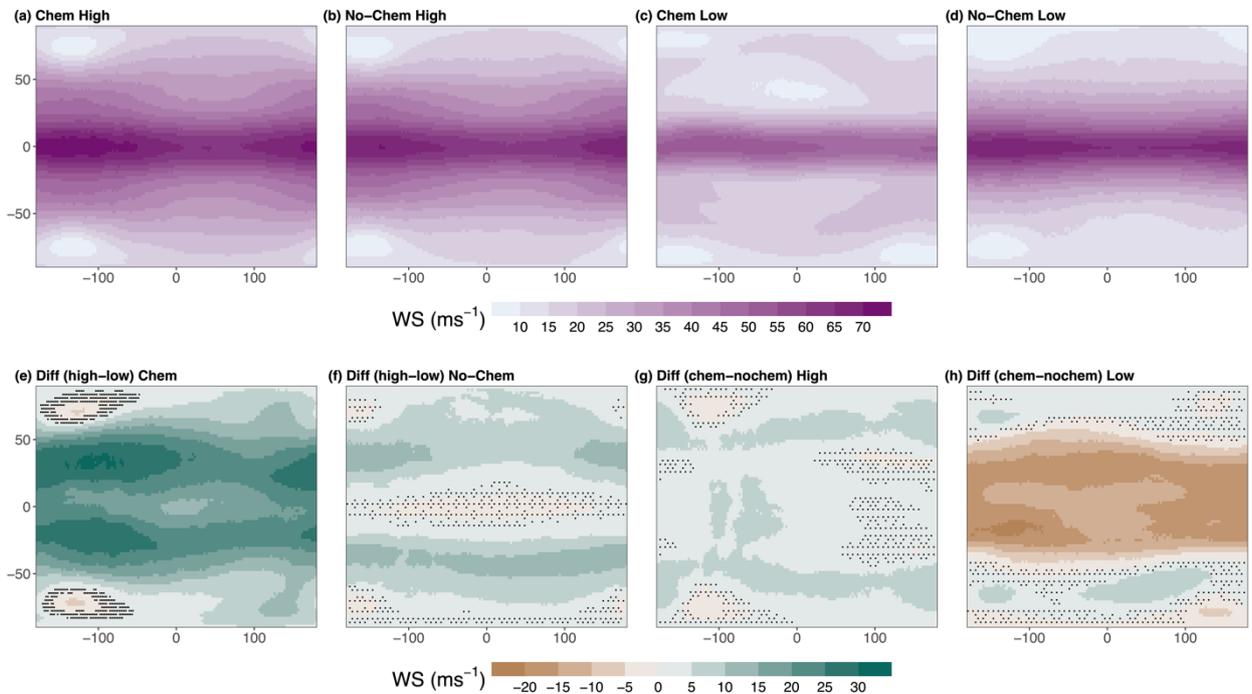

**Figure 6** - Same as Figure 5 but for Wind Speed (WS).

Similar to temperature and wind speed, we also computed the composite and difference maps for OzCol and OzFr, with the only difference being that here, we only use the Chemistry simulation. When looking at the Chemistry composite maps of both OzCol and OzFr during 'High' and 'Low' atmospheric states, we observe that higher states lead to higher accumulation of ozone over both the northern and southern gyres (Figure 7a-b, d-e), driven by a stratospheric dayside-to-nightside circulation (Braam et al., 2023). However, for OzFr, higher values of ozone are also found from 50°S to 90°S and from 50°N to 90°N, caused by the combined effect of atmospheric circulation and weaker chemical loss processes of ozone the further we move from the substellar point (Figure 7d) (e.g., Chen et al., 2018; Yates et al., 2020). Such spatial patterns are, therefore, reflected when looking at the difference maps of High - Low states. Indeed, we find positive differences for the 'High' states of OzCol over the northern and southern gyres, and the same is true for OzFr, but with more prominent enhanced values at higher latitudes. Moreover, the former variable shows negative differences over most of the remaining planet, and the latter shows negative and non-significant differences from ~10°S to ~10°N (Figure 7c, f). Hence, the composite maps of temperature, wind speed, and ozone together illustrate that the stratospheric dayside-to-nightside circulation that drives ozone accumulation over the gyres is most prevalent during the 'High' states. On the other hand, the 'Low' states represent a relatively weak dayside-to-nightside circulation, illustrated by less pronounced gyres and a more homogeneous ozone distribution.

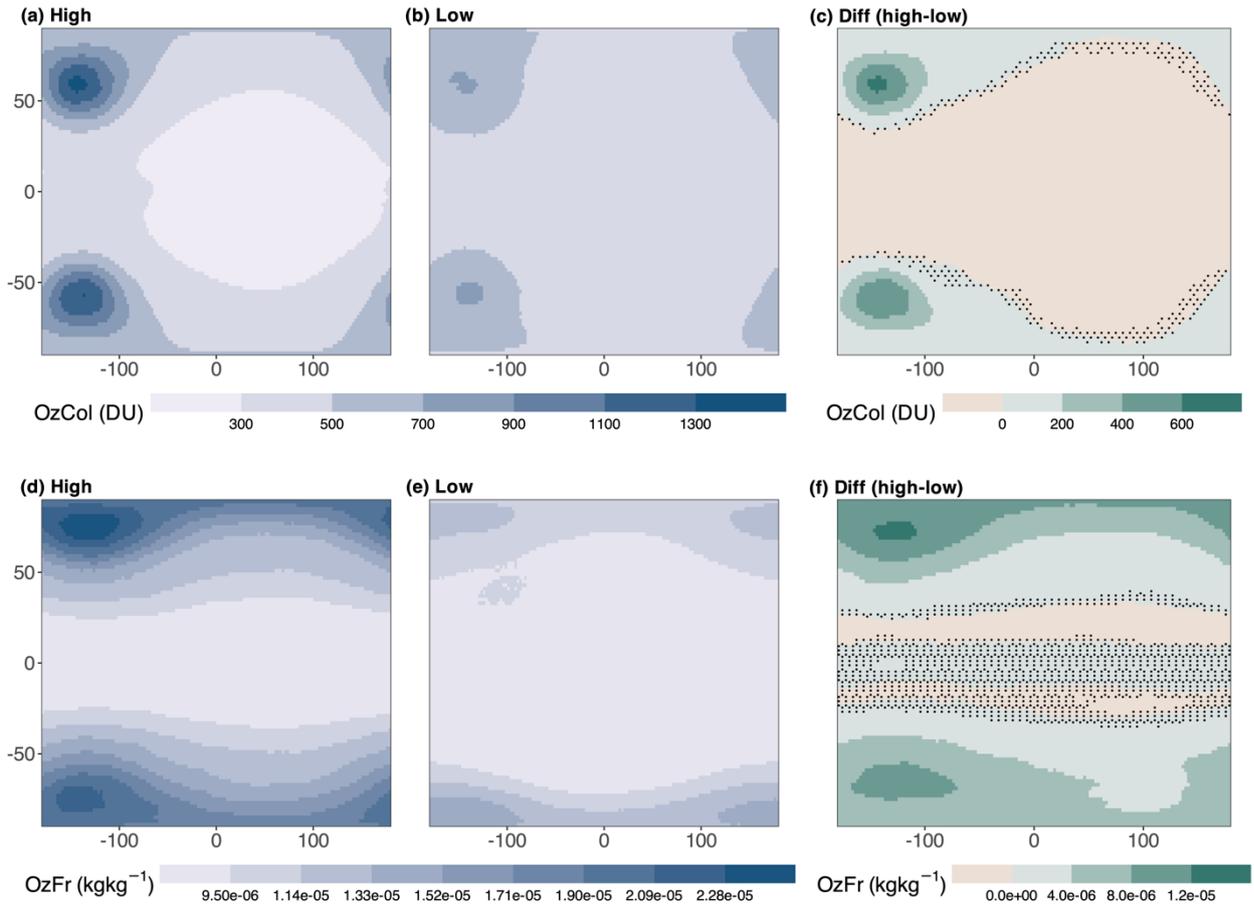

**Figure 7** - Same as Figure 5 but for OzCol (a-c) and OzFr (d-f). Note that the Figure shows the Chemistry simulation. In (c, f), stippling represents areas that are *not significantly* different at the 1% level.

## 4. Discussion and Conclusions

The atmospheres of tidally-locked terrestrial (or Earth-like) exoplanets are close to being characterized by the advent of the James Webb Space Telescope. This opens new frontiers in searching for habitable planets outside our solar system and incentivizes new ground-breaking investigations from the astrophysical, astronomical, and Earth sciences communities. With this work, we showed that adding an interactive ozone module to climate model simulations of Proxima Centauri b globally increases the stratospheric temperature and induces regionally varying effects on the surface temperature, including increased surface temperature in the gyre regions and a decrease of the dayside-to-nightside temperature contrast by 4 °K. Adding ozone in the simulations resulted in similar albeit significantly different median vertical compound dynamics of temperature and wind speed compared to not having interactive ozone, with more differences observed when assessing their standard deviations, indicating enhanced variability. We found that highly dynamically stable states have warmer gyre regions, stronger wind speeds, and enhanced ozone accumulation over the gyres.

Our findings can be summarized and contextualized as follows:

- **Chemistry model simulations show higher average stratospheric temperatures compared to No-Chemistry ones.** We attribute this difference to the ozone's radiative heating, a similar mechanism that we observe in the Earth's stratosphere due to the absorption and emission of terrestrial IR radiation and absorption of solar radiation in the UV and visible spectrums (Dopplick, 1972; Fishman et al., 1979; Park & London, 1974). The temperature difference was in the order of +7 °K and can be observed from ~17 km to ~50 km above the surface, the altitude range where the highest ozone mass fractions are found and where ozone is most likely to interact with the incoming stellar radiation. This agrees qualitatively with previous studies on the effect of ozone on vertical temperature structures, with Boutle et al. (2017) also reporting warming of the stratosphere for Proxima Centauri b. However, the quantitative warming effect of ozone varies. Godolt et al. (2015) showed that ozone heats the stratosphere of planets around F-type stars but does not significantly affect the vertical temperature structure of planets around K-type stars. The spectral dependence is further illustrated by Kozakis et al. (2022), showing that ozone abundances and the amount of stratospheric heating depend on the total amount of UV flux received from the host star and the distribution over wavelengths. The amount of incoming near-ultraviolet (NUV: $200 < \lambda < 400$ nm) radiation determines ozone production, while the far-ultraviolet (FUV: $91 < \lambda < 200$ nm) radiation determines the amount of oxygen photolysis and, thus, ozone production. Hence, the total UV flux and the FUV/NUV flux ratio are important metrics of ozone photochemistry and its impact on climate dynamics. Even if we only consider the photolysis wavelengths in UM-UKCA ($\lambda > 177$ nm), the MUSCLES spectrum used in this study had an FUV/NUV flux ratio of 0.012, a higher ratio than any of the host stars from Kozakis et al. (2022) and sufficient to drive mild stratospheric heating. The altitude that corresponds to the stratospheric ozone layer and its effect on the vertical temperature structure depends on the initial amount of $O_2$ present (Cooke et al., 2022, 2023).

- **The gyres of Proxima Centauri b show a "saddle" in the OzFr vertical profile, which the global ones do not.** Since the nightside is devoid of incoming radiation, the balance of ozone formation (three-body reaction $O + O_2 + M \rightarrow O_3 + M$) and ozone destruction (driven by photolysis and reaction with atomic O) in the Chapman mechanism shifts to more ozone production. This leads to a secondary nightside ozone layer at high altitudes. A mesospheric secondary ozone layer is also present during night-time on Earth (e.g., Smith & Marsh, 2005). Its permanent presence on the nightside of synchronously rotating planets was also found in simulations of Earth in a synchronous orbit (Proedrou & Hocke, 2016).

- **Adding interactive ozone to the simulations affects the compound dynamics of temperature and wind speed.** Both Chemistry and No-Chemistry simulations showed similar patterns of $\theta_{T, WS}$, and $d_{T,WS}$ vertical profiles when taking the medians over the 30 years in the entire planet, northern gyre, and western and eastern terminators. However, most of the vertical profiles showed statistically significant differences. We noticed more variability for the Chemistry simulation than the No-Chemistry one when considering the standard deviations. We therefore suggest that including interactive ozone improves the simulation of Proxima Centauri b's atmospheric time series dynamics, since increased variability broadly means a more realistic simulation of the atmospheric dynamics. Ozone, as a single interactive species, can significantly impact the compound dynamical metrics on both global and regional scales, depending on the stellar flux distribution and the initial oxygen content. From Figure 7 of Kozakis et al. (2022), ozone heats the atmosphere more for planets orbiting earlier stellar types. This relates to the FUV/NUV ratio, but perhaps even more so simply to the magnitude of the UV radiation, as shown in their Figure 2.

- **Highly dynamically stable states show larger ozone column and fraction over the gyres than 'Low' states.** 'Low' states reflect stalling (or more persistent) weather patterns compared to the more variable 'High' states. The Chemistry simulation showed a significant increase in OzCol in 'High' states over the northern and southern gyres' regions on the nightside, along with increased OzFr. On synchronously rotating planets, the day-nightside heating contrast generates an overturning circulation with rising air on the dayside and subsiding air on the nightside (Hammond & Lewis, 2021; Showman et al., 2013). The overturning circulation has a strong tropospheric component but extends into the stratosphere, with implications for photochemically generated ozone at these altitudes (Braam et al., 2023). These vertical motions, in turn, contribute to forming standing, planetary-scale Kelvin and Rossby waves, the latter of which manifest as the nightside gyres in our simulations (Showman et al., 2013). Photochemistry is not active in these non-irradiated gyre regions, and the enhanced OzCol is regulated by the stratospheric overturning circulation (Braam et al., 2023). This picture implies that the large-scale circulation, atmospheric variability, and OzCol variations are linked. We postulate that 'High' $\theta_{T,WS}$, and $d_{T,WS}$ states may represent migration in time for the gyres (see also Cohen et al., 2023), oscillating in their central longitude so that naturally, these states show enhanced variability. Furthermore, we postulate that these 'High' states have a particularly strong overturning circulation, enhancing the amount of ozone trapped in the gyres.

- **Wind speed for Chemistry 'Low' states is much lower than for 'High' states.** The existence of the gyres is related to the mechanism to form equatorial super-rotating jets on synchronous exoplanets since the planetary-scale waves (including the Rossby waves, which the gyres are lows of) pump eastward momentum equator-wards (Showman & Polvani, 2011). Given this mechanism to generate the jets, we suggest that the higher wind speed for 'High' states in our Chemistry simulation fits this complete dynamical picture and the finding of enhanced OzCol in the gyres for 'High' states. The simulations show pronounced and vigorous gyres and a strong equatorial jet for 'High' states. In contrast, the gyres are less pronounced in the 'Low' states, corresponding to less eastward momentum flowing equatorward, resulting in a substantially weaker jet.

Our results demonstrated the value of compound dynamical systems metrics to elucidate variability in the atmospheres of exoplanets. They can be extended beyond Proxima Centauri b to other Earth-like exoplanets. Our framework also has potential applications with future exoplanet observations, obtained, for example, by the James Webb Space Telescope, the Habitable Worlds Observatory, and the Large Interferometer for Exoplanets, since they will contribute to constraining the climate state, dynamics, and potential habitability of Earth-like exoplanets (Hochman et al., 2022, 2023; Quanz et al., 2022). Indeed, understanding how ozone impacts climate dynamics and its observations on exoplanets is crucial for grasping the potential habitability of distant worlds. Ozone plays a vital role in shielding an exoplanet from harmful UV radiation. The presence or absence of ozone can provide valuable insights into the composition and stability of exoplanets' atmospheres. By studying ozone and its interactions within different atmospheric environments, we can interpret atmospheric signatures observed in exoplanet atmospheres, helping to identify conditions conducive to life as we know it (Ben-Israel et al., 2024; Cole et al., 2020). Furthermore, understanding ozone dynamics aids in predicting how atmospheric changes, both natural and anthropogenic, may impact habitability on Earth and beyond, guiding our search for potentially habitable exoplanets in the vast universe. As a caveat, we acknowledge the fact that the actual radius of Proxima Centauri b is unknown because no transit has been detected so far (Kipping et al., 2017). Therefore, the planet may not be an Earth-like exoplanet (Brugger et al., 2017). We envisage future works on the impact of an entire interactive chemistry module on the climate dynamics of Earth-like exoplanets, with the case study being, for instance, Proxima Centauri b and TRAPPIST-1e. In addition, future work should also include a variety of host stars and FUV/NUV ratios and the effect of varying initial $O_2$ abundances.

## Acknowledgments

Paolo De Luca was funded by the European Union's Horizon Europe Research and Innovation Program under Grant Agreement 101059659. Marrick Braam is part of the CHAMELEON MC ITN EJD, which received funding from the European Union's Horizon 2020 research and innovation program under the Marie Sklodowska-Curie grant agreement no. 860470. The Israel Science Foundation (grant #978/23) funds the contribution of Assaf Hochman. We gratefully acknowledge using the MONSooN2 system, a collaborative facility supplied under the Joint Weather and Climate Research Programme, as a strategic partnership between the Met Office and the Natural Environment Research Council. The simulations were performed as part of the project space 'Using UKCA to investigate atmospheric composition on extra-solar planets (ExoChem)' with Principal Investigator Paul Palmer.

## Competing interests

The authors declare no competing interests.

## Data availability statement

The time series of the compound dynamical systems' metrics are available upon reasonable request to the corresponding author.